\documentclass[11pt]{article}
\usepackage{geometry}
\usepackage{titlesec}
\usepackage[colorlinks,citecolor=blue,linkcolor=blue,urlcolor=blue]{hyperref}
\usepackage{graphicx}
\usepackage{fancyhdr}
\usepackage{amsmath, amssymb}
\usepackage{enumitem}
\usepackage{microtype}
\usepackage{orcidlink}
\usepackage{tikz}
\usetikzlibrary{positioning, shapes.multipart, shapes.geometric, arrows.meta}
\usepackage{xcolor}

\title{\textbf{Single Conversation Methodology: A Human-Centered Protocol for AI-Assisted Software Development}}
\author{Salvador D. Escobedo\,\orcidlink{0000-0002-1545-2080} \footnote{\small salvadec@live.com.mx, sdcasillas@stefanini.com}}
\date{\today}

\begin{document}
\maketitle

\begin{abstract}
We propose the \emph{Single Conversation Methodology} (SCM), a novel and pragmatic approach to software development using large language models (LLMs). In contrast to ad hoc interactions with generative AI, SCM emphasizes a structured and persistent development dialogue, where all stages of a project—from requirements to architecture and implementation—unfold within a single, long-context conversation. The methodology is grounded on principles of cognitive clarity, traceability, modularity, and documentation. We define its phases, best practices, and philosophical stance, while arguing that SCM offers a necessary correction to the passive reliance on LLMs prevalent in current practices. We aim to reassert the active role of the developer as architect and supervisor of the intelligent tool.
\end{abstract}


\section{Introduction}

The integration of large language models (LLMs) into software development has introduced significant changes in how programmers interact with code \cite{haque2024llms, jalil2023transformative}. These models offer unprecedented capabilities in code generation, refactoring, and debugging, reshaping not only development workflows but also the cognitive practices of developers \cite{tabarsi2025reshaping, zhang2024survey}. However, their ease of use and on-demand nature have also led to engagement patterns that can undermine key aspects of sound software engineering \cite{khojah2024beyond, edwards2025vibes}.

A dominant pattern has emerged: developers engage in short, goal-oriented prompts with minimal contextual continuity. A typical session consists of requesting a function, copying the output, and terminating the interaction. This piecemeal approach, while expedient, promotes what we may call a \textit{fragmented development mindset}. Code is accepted based on surface-level utility rather than on architectural soundness, long-term maintainability, or conceptual clarity \cite{hou2024large, khojah2024beyond}.

This trend echoes what has been described as \emph{vibe coding}—a style of programming characterized by intuition, speed, and exploratory trial-and-error rather than deliberate design \cite{chowdhury2025vibecoding, roose2025notacoder, edwards2025vibes}. While useful in early prototyping or creative experimentation, vibe coding becomes problematic when applied to complex systems or production-grade software. Its informal, improvisational nature often leads to opaque logic, weak modular boundaries, and fragile interdependencies. When LLMs are used in this mode, they tend to amplify these weaknesses by encouraging rapid iteration without reflection.

The central risk is not merely technical debt, but cognitive erosion. As developers outsource more of the planning and reasoning to the model, they may lose their grip on the architectural vision of the project \cite{atemkeng2024ethics, kazemitabaar2025cognitive}. Over time, this can result in software systems that are difficult to reason about, evolve, or even explain.

The \emph{Single Conversation Methodology} (SCM) was conceived as a principled alternative to such practices. Rather than reducing the human developer to a passive prompter, SCM proposes a development protocol in which the conversation with the LLM is treated as a coherent, structured, and persistent workspace. The developer remains in full control of the design direction, using the model as a context-rich assistant rather than a disposable code generator.

SCM restores a sense of continuity and intentionality to AI-assisted development. It encourages disciplined reasoning, architectural awareness, and modular design through a methodical use of dialogue. In this way, it addresses both the practical shortcomings of ad hoc prompting and the philosophical concern of maintaining human agency in an increasingly automated workflow \cite{atemkeng2024ethics, wang2024trust}.

In practice, the Single Conversation Methodology has been successfully applied using Claude 3.7, whose extended context window enables coherent long-form development sessions. However, the methodology is model-agnostic in principle: any sufficiently capable language model with a large enough context window to maintain architectural continuity and recall prior interactions can support SCM. The key requirement is not a specific model, but the ability to sustain a persistent, structured dialogue where design reasoning, implementation steps, and revisions remain accessible throughout the project lifecycle.

\section{Methodology Overview}

The Single Conversation Methodology (SCM) reimagines the process of software development in the age of large language models by treating the LLM interaction itself as the primary development environment. Unlike conventional workflows fragmented across multiple tools and prompts, SCM unfolds entirely within a single, persistent conversation thread. This conversation becomes the developer's workspace, memory, and reasoning context. The model assists not merely as a code emitter but as a high-context collaborator throughout the lifecycle of the project.

\subsection{Core Principle}

\begin{quote}
\emph{The conversation is the development environment. The developer remains the architect; the LLM serves as a high-context assistant.}
\end{quote}

This principle underlines a clear division of roles: the human retains strategic control, while the model contributes local insights, structured outputs, and memory continuity. The use of conversational programming interfaces as assistant personas has been explored in recent work~\cite{ross2023conversational}, validating the idea that dialogic interaction enables richer forms of collaboration than traditional code generation. SCM restores intentionality and coherence to the development process by grounding every action within a shared, evolving context. At the same time, critical oversight remains necessary to ensure correctness and trust~\cite{wang2024trust}.

\subsection{Phases of the Methodology}

SCM unfolds through a structured sequence of phases, each governed by explicit transitions and all anchored in the same conversational thread. This design ensures that architectural decisions, rationales, and implementation history are preserved within a unified dialogue.

\begin{figure}[ht]
\centering
\begin{tikzpicture}

\def\blockwidth{5.2}

\draw[fill=blue!20,draw=black] (0,0) rectangle (\blockwidth,-1.0);
\node[align=left] at (2.6,-0.5) {\small{Grounding phase}};

\draw[fill=gray!20,draw=black] (0,-1.0) rectangle (\blockwidth,-4.8);
\node[align=left] at (2.6,-2.8) {\small{Code Generation}};

\draw[fill=brown!20,draw=black] (0,-4.8) rectangle (\blockwidth,-5.8);
\node[align=left] at (2.6,-5.3) {\small{Documentation}};

\node[above] at (3,0.5) {\textbf{General SCM structure}};
\draw[->, thick, dashed] (\blockwidth + 0.5, 0) -- (\blockwidth + 0.5, -6.1);
\node[rotate=270] at (\blockwidth + 1.0, -3) {\textit{Temporal Flow of the Conversation}};

\end{tikzpicture}
\caption{The SCM has three main stages: Grounding phase, Code Generation and Documentation.}
\label{fig:overview}
\end{figure}
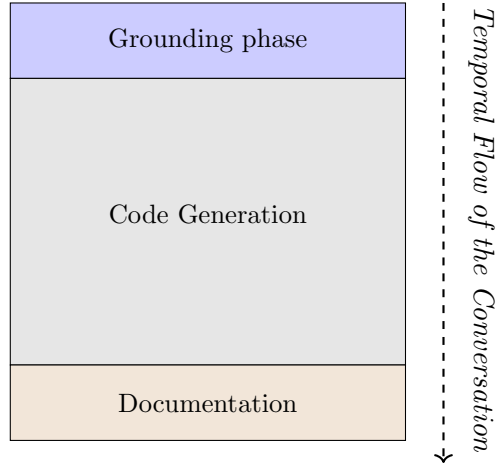

The overall structure recommended by SCM has three stages: (1) The \textit{grounding phase}, where the generalities of the project and setup are discussed and defined. No code is generated in this phase. (2) The \textit{code generation} phase, where the code is actually generated, following certain cyclic patterns, and (3) the \textit{documentation} phase, where design concepts and project general documentation artifacts are generated. (see figure \ref{fig:overview})

Code generation phase usually takes the largest part of the dialogue, but grounding phase can also be extensive; it depends on the project and the capacity of the model window.

\subsubsection{Grounding Phase (No Code)}

In this first stage, the project begins with a conceptual grounding discussion. No code is written during this phase. Instead, the goal is to establish a mutual understanding between the developer and the model.

\begin{itemize}
    \item Define functional and non-functional requirements.
    \item Describe the overall architectural vision and technological stack.
    \item Clarify domain boundaries, design patterns, and architectural constraints.
    \item Align expectations and clarify terminology to avoid downstream ambiguity.
\end{itemize}

Only after a coherent conceptual base is established should the development of actual code begin. This avoids premature generation and ensures that subsequent outputs are contextually grounded. Recent studies highlight the advantages of long-context interaction with LLMs~\cite{zhang2024survey}, confirming the value of sustained, coherent dialogue in complex tasks such as architectural planning.

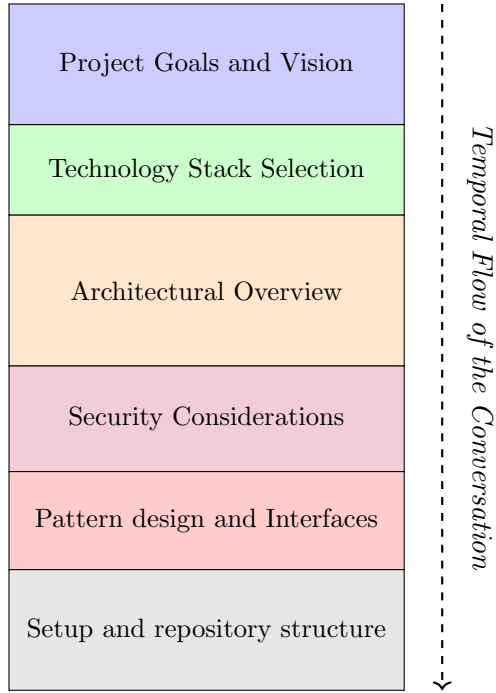
\begin{figure}[!ht]
\centering
\begin{tikzpicture}

\def\blockwidth{5.2}

\draw[fill=blue!20,draw=black] (0,0) rectangle (\blockwidth,-1.6);
\node[align=left] at (2.6,-0.8) {\small{Project Goals and Vision}};

\draw[fill=green!20,draw=black] (0,-1.6) rectangle (\blockwidth,-2.8);
\node[align=left] at (2.6,-2.2) {\small{Technology Stack Selection}};

\draw[fill=orange!20,draw=black] (0,-2.8) rectangle (\blockwidth,-4.8);
\node[align=left] at (2.6,-3.8) {\small{Architectural Overview}};

\draw[fill=purple!20,draw=black] (0,-4.8) rectangle (\blockwidth,-6.2);
\node[align=left] at (2.6,-5.5) {\small{Security Considerations}};

\draw[fill=red!20,draw=black] (0,-6.2) rectangle (\blockwidth,-7.5);
\node[align=left] at (2.6,-6.85) {\small{Pattern design and Interfaces}};

\draw[fill=gray!20,draw=black] (0,-7.5) rectangle (\blockwidth,-9.1);
\node[align=left] at (2.6,-8.3) {\small{Setup and repository structure}};

\node[above] at (3,0.5) {\textbf{Grounding Phase Conversation Overview}};
\draw[->, thick, dashed] (\blockwidth + 0.5, 0) -- (\blockwidth + 0.5, -9.1);
\node[rotate=270] at (\blockwidth + 1.0, -4.55) {\textit{Temporal Flow of the Conversation}};

\end{tikzpicture}
\caption{Illustration of distinct thematic blocks in an LLM-based Grounding Phase. Each section corresponds to a focused design discussion.}
\label{fig:grounding}
\end{figure}

The schematic conversation in the figure \ref{fig:grounding} exemplifies the purpose of the Grounding Phase: establishing shared vocabulary, validating requirements, and framing the architectural vision—without writing a single line of code. Only after this understanding is reached does the methodology proceed to modular development. There is no fixed rule for structuring this grounding discussion, and the programmer should be able to develop an adequate grounding schema for the specific needs of the project. 

\subsubsection{Code Generation Stage}

\begin{figure}[ht]
\centering
\begin{tikzpicture}

\def\blockwidth{5.2}

\draw[fill=blue!20,draw=black] (0,0) rectangle (\blockwidth,-1.6);
\node[align=left] at (2.6,-0.8) {\small{Analysis}};

\draw[fill=green!20,draw=black] (0,-1.6) rectangle (\blockwidth,-2.8);
\node[align=left] at (2.6,-2.2) {\small{Code Generation}};

\draw[fill=orange!20,draw=black] (0,-2.8) rectangle (\blockwidth,-4.8);
\node[align=left] at (2.6,-3.8) {\small{Troubleshooting}};

\draw[fill=purple!20,draw=black] (0,-4.8) rectangle (\blockwidth,-6.2);
\node[align=left] at (2.6,-5.5) {\small{Summary}};

\node[above] at (3,0.5) {\textbf{Modular Development Cycle}};
\draw[->, thick, dashed] (\blockwidth + 0.5, 0) -- (\blockwidth + 0.5, -6.1);
\node[rotate=270] at (\blockwidth + 1.0, -3) {\textit{Temporal Flow of the Conversation}};

\end{tikzpicture}
\caption{Illustration of distinct thematic blocks in an LLM-based Modular Development Cycle. Each section corresponds to a focused development process of a specific layer or module.}
\label{fig:develop}
\end{figure}
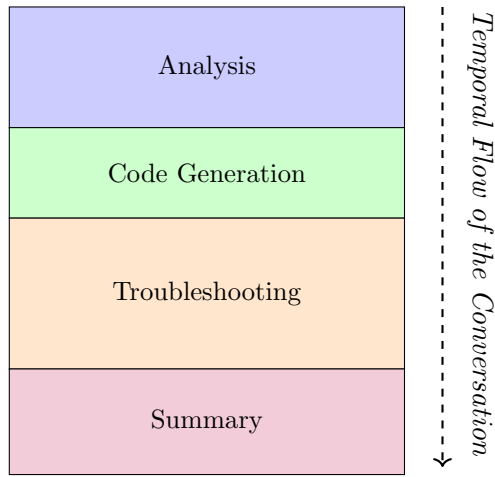

A second general stage in the SCM is devoted to code generation. SCM promotes modularity at both the architectural and conversational levels, dividing this part of the interaction into a number of conversation cycles. Each component or subsystem of the architecture is developed through a structured conversational cycle, maintaining local coherence while contributing to the global system. Each cycle is divided as follows (see figure \ref{fig:develop}):

\paragraph{A. Analysis.} Discuss the purpose, responsibilities, interfaces, dependencies, and placement of the component within the overall architecture.

\paragraph{B. Code Generation.} Request code in small, logically grouped files or units. Avoid large, undifferentiated code dumps. Favor clarity, naming consistency, and testability.

\paragraph{C. Troubleshooting.} Address only code-related errors and logical issues. Broader concerns such as tooling or environment setup should be managed outside the LLM context.

\paragraph{D. Summary.} Recapitulate what has been implemented, how it fits into the system, and any insights or decisions that arose. This reinforces conceptual continuity and supports future review.

\subsubsection{Iterative Expansion}

As the project evolves, new components, features, or concerns may be introduced. SCM emphasizes clear transitions between topics of the conversation. For example:

\begin{quote}
\emph{“Let’s now proceed to the API gateway layer.”}
\end{quote}

These explicit transitions serve multiple purposes: they structure the dialogue, allow future navigation and traceability, and prepare the conversation for being reused or resumed at later stages. Each addition builds upon prior context without fragmenting the workflow.

\subsubsection{Documentation Generation}

In SCM, documentation stage is not treated as a separate post-development obligation, but as an organic continuation of the same high-context conversation that produced the implementation. However, we recommend deferring the creation of formal documentation artifacts until the end of the project, once the architecture, naming conventions, and component interfaces have stabilized.

\medskip

At that stage, the developer can prompt the LLM to generate cohesive documentation based on the shared conversational history. This typically includes:

\begin{itemize}
    \item Architectural overviews and system diagrams.
    \item Design rationales for key modules, classes, and subsystems.
    \item Interface specifications with semantic and structural annotations.
    \item Onboarding guides and usage instructions tailored to new developers.
\end{itemize}

Because the LLM retains full awareness of the development process, the resulting documentation is internally consistent, contextually accurate, and often surprisingly complete~\cite{chakrabarty2025documentation, macke2024documentation}.

\medskip

Throughout development, the LLM can also assist in producing informal summaries, quick rationales, or snapshots of component structure. These conversational notes serve as scaffolding for later synthesis. In particular, SCM enables the automatic generation of high-quality pull request (PR) descriptions, grounded in prior conversation and development rationale. This is especially valuable during code review phases, as it enhances reviewer understanding, preserves design intent, and improves the project's traceability over time.

\medskip

This approach combines the benefits of continuous narrative memory with the clarity of post hoc synthesis, reducing the burden of documentation while improving its quality and relevance.

\pagebreak
\section{Best Practices}

SCM includes a number of guiding principles that enhance its clarity and power:

\begin{enumerate}
    \item \textbf{Structure by Layers or Modules.} Avoid working at the file level without architectural context.
    \item \textbf{Defer Code.} Do not generate code until the model has shown understanding of the requirements and design.
    \item \textbf{Stay on Topic.} Avoid troubleshooting issues not related to the current software (e.g., unrelated environment errors).
    \item \textbf{Mark Transitions.} Clearly indicate topic changes to maintain structure.
    \item \textbf{Generate in Small Batches.} Avoid overwhelming output and maintain clarity.
    \item \textbf{Use the Model for Design Documents.} Ask the model to generate human-readable summaries of design decisions.
\end{enumerate}
\section{Philosophical Foundation}

At its heart, SCM expresses a clear stance on human–machine collaboration:

\begin{quote}
\emph{The programmer must not follow the AI. The AI must follow the programmer.}
\end{quote}

This principle safeguards the intellectual integrity of the engineering process. It asserts that software development must remain a human-directed activity, where powerful tools—no matter how sophisticated—serve the vision, intention, and judgment of their users.

SCM is not merely a method for integrating LLMs into development workflows. It is a practical philosophy: one that affirms the centrality of human agency in the creative act of engineering. Its key commitments include:

\begin{itemize}
  \item \textbf{Epistemic Responsibility.} The developer must understand what is being built. SCM requires that all architectural decisions and technical implementations remain explainable and intelligible to the human author.
  
  \item \textbf{Critical Oversight.} AI outputs are treated as suggestions—not authorities. SCM encourages an active, questioning stance: every design proposal or code snippet must pass through the developer’s scrutiny before being accepted.
  
  \item \textbf{Intentional Craft.} Software development is not a passive activity. The programmer is not a curator of AI-generated artifacts, but a designer engaged in a deliberate, goal-driven process. The LLM serves to accelerate this process—not to steer it.
  
  \item \textbf{Traceable Intent.} SCM preserves context across time. Because all decisions unfold in a single, persistent conversation, future maintainers can recover not only what was built, but why. This fosters transparency, continuity, and resilience.
\end{itemize}

The Single Conversation Methodology strikes a balance between two extremes: on one side, the undisciplined enthusiasm of freeform code generation; on the other, the rigidity of heavyweight methodologies. SCM offers a lightweight yet structured alternative—one that:

\begin{itemize}
  \item Harnesses the generative power of LLMs to streamline routine work.
  \item Preserves the developer's role as the source of architectural vision and critical judgment.
  \item Maintains coherence through a single, high-context development thread.
  \item Supports rapid iteration without compromising design clarity or long-term maintainability.
\end{itemize}

In this way, SCM is not just a technique—it is a defense of responsible, creative, and human-centered software engineering in the age of artificial intelligence.

\section{Advanced Applications of SCM}

While SCM is particularly intuitive for greenfield projects, it is equally applicable to complex and collaborative development scenarios, including those involving existing codebases and distributed teams. In these contexts, the method retains its clarity and consistency through strategic use of Retrieval-Augmented Generation (RAG) and disciplined conversation management.

\subsection{Development with Existing Codebases}

In mature or legacy projects, development often begins not from scratch, but from the need to extend, refactor, or fix existing code. SCM adapts seamlessly to this reality by leveraging RAG to provide the LLM with contextual access to the codebase \cite{koziolek2024rag}. In this setup, the conversation begins not with architectural grounding, but with a feature request or bug report.

\medskip

\textbf{Workflow Overview:}
\begin{enumerate}
    \item The developer begins by stating the feature or fix to be implemented.
    \item The LLM, powered by RAG, inspects relevant parts of the codebase.
    \item A high-context discussion ensues about the design rationale, interfaces, and potential side effects.
    \item Modular implementation proceeds using the same SCM sub-cycle: \emph{analysis, code generation, troubleshooting, summary}.
    \item Documentation and PR descriptions are synthesized at the end of the session.
\end{enumerate}

This approach allows developers to integrate seamlessly into unfamiliar or large systems while retaining a clear and traceable development narrative. It has been shown to significantly benefit specialized developer groups by improving access to contextual knowledge and reducing onboarding time \cite{flores2025visually}.

\subsection{Collaborative Development and Parallel Conversations}

SCM also supports multi-developer environments by encouraging each programmer to maintain their own persistent conversation with the LLM. Because the model accesses the shared codebase via RAG, each session can operate independently while staying contextually grounded \cite{du2024multiagent}.

\medskip

\textbf{Guidelines for Collaboration:}
\begin{itemize}
    \item Each developer should scope their work clearly, beginning their conversation with a definition of goals and architectural boundaries.
    \item Frequent synchronization should occur at integration points, typically mediated by version control systems and coordinated through shared documentation or pull request reviews.
    \item Developers can use shared documentation prompts or templates to ensure consistency across LLM-generated artifacts.
\end{itemize}

This structure enables high parallelism without sacrificing coherence. Importantly, SCM does not require the LLM to maintain a global memory of all conversations—RAG serves as the connective tissue, grounding each local session in a consistent and shared reality.

\medskip

By combining persistent, goal-oriented sessions with global codebase awareness, SCM enables distributed AI-assisted development that is modular, traceable, and collaborative by design.

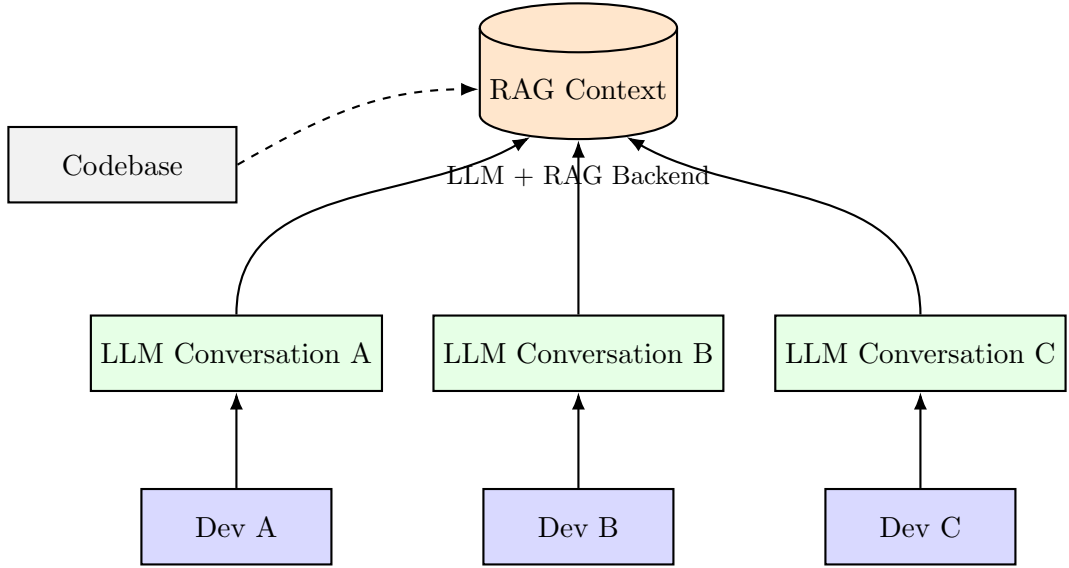
\begin{figure}[h]
\centering
\begin{tikzpicture}[
  node distance=1.2cm and 2.5cm,
  dev/.style={rectangle, draw, thick, fill=blue!15, minimum width=2.5cm, minimum height=1cm},
  conv/.style={rectangle, draw, thick, fill=green!10, minimum width=3.5cm, minimum height=1cm},
  rag/.style={cylinder, shape border rotate=90, aspect=0.25, draw, thick, minimum height=1.8cm, minimum width=1.2cm, fill=orange!20},
  code/.style={rectangle, draw, thick, fill=gray!10, minimum width=3cm, minimum height=1cm},
  arrow/.style={thick, -{Latex[]}}
]

\node[rag] (rag) at (0,4) {RAG Context};

\node[code] (codebase) at (-6,3) {Codebase};

\node[dev] (dev1) at (-4.5,-1.8) {Dev A};
\node[conv] (conv1) at (-4.5,0.5) {LLM Conversation A};

\node[dev] (dev2) at (0,-1.8) {Dev B};
\node[conv] (conv2) at (0,0.5) {LLM Conversation B};

\node[dev] (dev3) at (4.5,-1.8) {Dev C};
\node[conv] (conv3) at (4.5,0.5) {LLM Conversation C};

\draw[arrow] (dev1) -- (conv1);
\draw[arrow] (dev2) -- (conv2);
\draw[arrow] (dev3) -- (conv3);

\draw[arrow] (conv1.north) to[out=90,in=210] (rag.south west);
\draw[arrow] (conv2.north) -- (rag.south);
\draw[arrow] (conv3.north) to[out=90,in=330] (rag.south east);

\draw[arrow, dashed] (codebase.east) to[out=30,in=180] (rag.west) node[midway, above, sloped] {};

\node[below=0.2cm of rag] {\small{LLM + RAG Backend}};

\end{tikzpicture}
\caption{Collaborative SCM with Shared RAG Access to Existing Codebase}
\end{figure}

\section{Conclusion}

In an era where AI-driven code generation promises unprecedented speed and convenience, the Single Conversation Methodology offers a principled counterbalance. By anchoring every design decision, implementation step, and documentation artifact within one sustained dialogue, SCM preserves the developer’s authority and expertise. It leverages the strengths of large language models—context awareness, rapid synthesis, and creative suggestion—while safeguarding against uncritical acceptance and architectural drift. 

Through its lightweight yet disciplined structure, SCM transforms the LLM interaction from a series of ad hoc prompts into a coherent development workflow. Developers remain actively engaged as architects, critics, and narrators of their own projects, harnessing AI as an assistant rather than relinquishing control. As a result, teams can move quickly without sacrificing clarity, maintainability, or collective understanding. 

Looking ahead, the principles of SCM may be extended and adapted to new modalities—multimodal models, tighter IDE integrations \cite{du2024multiagent,koziolek2024rag}, or even fully automated code review assistants—always guided by the fundamental maxim: the programmer does not follow the AI; the AI follows the programmer. By embedding this philosophy into everyday practice, we can ensure that the promise of AI in software engineering is realized responsibly, creatively, and sustainably.

\bibliographystyle{plain}
\bibliography{references}


\end{document}